\magnification=\magstep1
\centerline{\bf Obtaining the Spin of the Proton from the Spins of the
Quarks}
\bigskip
\centerline{Y.N. Srivastava and A. Widom}
\centerline{Physics Department, Northeastern University, Boston MA 02115}
\centerline{and}
\centerline{Physics Department \& INFN, University of Perugia, Perugia
Italy}
\vskip 1cm
\centerline{ABSTRACT}
\medskip 

{\it Employing the experimental electro-weak interactions between quarks 
and leptons, the spin of the proton may easily be understood in terms of 
the spins of the constituent quarks.} 
\vskip 1cm

Before considering how one may obtain the spin of the proton 
from the spin of the constituitive qurks,  let us recall a simple 
fact about total spin zero momentum space wave functions for any two 
spin one half particles; e.g.  
$$
\Psi_{spin\ zero}({\bf k}_1,\sigma_1 ,{\bf k}_2,\sigma_2)=\sqrt{1\over 2}
\ \Phi ({\bf k}_1,{\bf k}_2)
\left(\chi_{\uparrow }(\sigma_1)\chi_{\downarrow }(\sigma_2)-
\chi_{\uparrow }(\sigma_2)\chi_{\downarrow }(\sigma_1)\right).
\eqno(1)
$$
The probability distribution in momentum and spin for one of 
the particles, for example  
$$
P_1({\bf k}_1,\sigma_1) =\sum_{\sigma_2}
\ \int {d^3 {\bf k}_2\over (2\pi )^3}
\left|\Psi({\bf k}_1,\sigma_1 ,{\bf k}_2,\sigma_2)\right|^2, 
\eqno(2)
$$
does not at all depend on spin 
$$
P_1({\bf k}_1,\uparrow )=P_1({\bf k}_1,\downarrow )
={1\over 2}\int {d^3 {\bf k}_2\over (2\pi )^3}
\left|\Phi ({\bf k}_1,{\bf k}_2)\right|^2. 
\eqno(3)
$$
Such an equipartition of spin probability would certainly 
{\it not} hold true for a total spin one polarized wave function; e.g.  
$\Psi_{spin\ one}({\bf k}_1,\sigma_1 ,{\bf k}_2,\sigma_2) 
=\Phi_{triplet}({\bf k}_1,{\bf k}_2)
\chi_\uparrow (\sigma_1)\chi_\uparrow (\sigma_2)$.

Now, suppose that the proton is made up of three quarks $(duu)$, and 
further suppose the quarks are distributed as a spin zero diquark [1,2] 
and a spin one half quark. Say 
$$
(p)=(duu)=(du)_{(spinless\ diquark)}+u_{(quark)} 
$$
tied together by a ``string'' as in conventional fragmentation models.
Under the above hypothesis, the  momentum probabilities are independent 
of spin for the quarks in the diquark, {\it both $d\in (du)$ 
and $u\in (du)$}, and only the isolated quark $(u)$ may have a 
momentum probability which depends on polarization. 
From lepton-proton scattering polarization measurements of spin, 
$$
s_{polarized}=\left({1\over 2}\right)\eta , \eqno(3)
$$
{\it only one third of the constituent quarks} would theoretically 
exhibit the polarization in their momentum distributions, 
$$
\eta_{theory} =\left({1\over 3}\right)\ne 1\ . \eqno(4)
$$
The experimental number is[3-7] 
$$ 
\eta_{experiment}=0.31\pm 0.07 \eqno(5)
$$ 
in very reasonable agreement with the physical picture here presented.
\vskip 1cm
\centerline{REFERENCES}
\vskip .5cm
\par \noindent
[1] For a general review on the subject of diquarks, see
M. Anselmino, E. Predazzi, S. Ekelin, S. Frederikson and D. B. 
Lichtenberg, {\it Rev. Mod. Phys.} {\bf 65} (1993) 1199.
\par\noindent
[2] For supersymmetry related to diquarks, see A. B. Balantekin,
I. Bars and F. Iachello, {\it Phys. Rev. Lett.} {\bf 47} (1981) 19;
{\it Nuc. Phys.} {\bf A370} (1981) 284.
\par\noindent
[3] R. K. Ellis, W. J. Stirling and B. R. Webber, ``QCD and Collider
Physics'', chapter 4, page 150 Eq.(4.243),
Cambridge Monographs on Particle Physics, Nuclear
Physics and Cosmology, Cambridge University Press UK (1996). 
\par\noindent
[4] M. Algard {\it et al}, {\it Phys. Rev. Lett.} {\bf 37}
(1976) 1261; G. Baum {\it et al}, {\it Phys. Rev. Lett.} {\bf 45}
(1980) 2000; ibid. {\bf 51} (1983) 1135. [SLAC-E130 collaboration].
\par\noindent
[5] K. Abe {\it et al}, {\it Phys. Rev. Lett.} {\bf 74} (1995) 346;
ibid. {\bf 75} (1995) 25. [SLAC-E143 collaboration]
\par\noindent
[6] J. Ashman {\it et al}, {\it Nuc. Phys.} {\bf B328} (1989) 1. 
[EMC collaboration]
\par\noindent
[7] B. Adeva {\it et al}, {\it Phys. Lett.} {\bf B302} (1993) 533;
D. Adams {\it et al}, {\it Phys. Lett.} {\bf B329} (1994) 399;
ibid. {\bf B357} (1995) 248. 
[SMC collaboration]

\bye